\documentclass[amssymb,amsmath,pra,twocolumn,floatfix,showpacs]{revtex4}
\usepackage{graphicx}
\usepackage{lastpage}
\usepackage[normalem]{ulem}

\newcommand{\ket}[1]{|#1 \rangle}
\newcommand{\bra}[1]{\langle #1|}
\newcommand{\braket}[2]{\langle #1| #2\rangle}

\begin{document}
\title{Dark State Adiabatic Passage with spin-one particles}

\author{Andrew D.~Greentree}
\affiliation{Chemical and Quantum Physics, School of Applied Sciences, RMIT University, Melbourne 3001, Australia}
\email{andrew.greentree@rmit.edu.au}

\author{Belita Koiller}
\affiliation{Instituto de F\'{i}sica, Universidade Federal do Rio de Janeiro, Cx.Postal 68528, RJ 21941-972, Brazil}
\email{bk@if.ufrj.br}

\date{\today}

\begin{abstract}
Adiabatic transport of information is a widely invoked resource in connection with quantum information processing and distribution. The study of adiabatic transport via spin-half chains or clusters is standard in the literature, while in practice the true realisation of a completely isolated two-level quantum system is not achievable. We explore here, theoretically, the extension of spin-half chain models to higher spins. Considering arrangements of three spin-one particles, we show that adiabatic transport, specifically a generalisation of the Dark State Adiabatic Passage procedure, is applicable to spin-one systems. We thus demonstrate a qutrit state transfer protocol. We discuss possible ways to physically implement this protocol, considering quantum dot and nitrogen-vacancy implementations.
\end{abstract}

\pacs{03.67.Hk, 05.60.Gg, 75.10.Pq}

\maketitle

\section{Introduction}
The communication of information around small quantum networks is becoming increasingly important as control and design of such quantum systems becomes more advanced.  There are now many different approaches to such transport \cite{bib:Bose2007} and the choice of the `best' protocol for a given task depends on the size of the quantum system and the level of control that can be applied to it.

One class of transport protocols that is of interest is the set of protocols inspired by adiabatic passage.  Generically, adiabatic passage is the controlled evolution of a quantum system from an initial to a final state, so as to maintain the system in an instantaneous eigenstate throughout, by means of control of both tunnel matrix elements and on-site energies.  The canonical example of adiabatic passage is perhaps STIRAP, STImulated Raman Adiabatic Passage \cite{bib:GRB+1988} (see Refs.~\cite{bib:VitanovARPC2001,bib:KTS2007} for a good discussion of this, and related adiabatic techniques).  Here, an excitation (typically an electron) is moved between energy levels in a three or more level atomic system.  The only control is via coherent electro-magnetic fields (e.g. lasers) and the so-called Counter-Intuitive pulse sequence (defined below) is employed.

Although many extensions of STIRAP are possible, in general the natural restrictions of using atomic systems can limit what is possible or practical.  However combining STIRAP techniques with spatially engineered systems mitigates this restriction somewhat, as seen in original work applying STIRAP techniques to double quantum dot systems \cite{bib:BrandesPRL2000,bib:BrandesPRB2001}.  Later, full spatial variants of STIRAP were explored including the Coherent Tunneling Adiabatic Passage (CTAP) approach, which has been studied in the context of atoms in triple well potentials \cite{bib:EckertPRA2004}, superconductors \cite{bib:Siewert2004}, electrons bound to quantum dots and to donors \cite{bib:GreentreePRB2004}, Bose-Einstein condensates \cite{bib:GraefePRA2006,bib:RabPRA2008}, photons in waveguides \cite{bib:Paspalakis2006,bib:Longhi2007} and Bose-Hubbard systems \cite{bib:BradlyPRA2012}.  Again, the main strength of CTAP derives from the ability to engineer the Hilbert space for certain functions, and in this context there exist applications for quantum information transport \cite{bib:GreentreePRB2004,bib:HollenbergPRB2006}, adiabatic splitting and operator measurements \cite{bib:GreentreePRA2006,bib:DevittQIP2007}, quantum gates \cite{bib:KestnerPRA2011}, interferometry \cite{bib:JongPRB2010,bib:Rab2012}, and branching structures for interaction-free measurement \cite{bib:HillNJP2011} and multi-port splitting \cite{bib:RangelovPRA2012}.  Most generally, adiabatic passage techniques can be understood as implementing generalised Morris-Shore transformations \cite{bib:MS1983,bib:RVS2006}.

Another scheme related to STIRAP that also takes advantage of Hilbert space engineering is Dark State Adiabatic Passage, DSAP \cite{bib:Ohshima2007,bib:OSF+2013}. The dark state in a three-level $\Lambda$ system is an eigenstate, which has no overlap with the excited state, and is the eigenstate that is utilised by the STIRAP process.  DSAP is named for the multi-spin generalisation of this state, although technically, the term dark state is not meaningful in DSAP as there is no requirement for an optically active excited state to be present in the system. In DSAP, a spin chain is considered with adiabatically controlled spin-spin couplings.  Formally, if the chain is a one-dimensional array of spin 1/2 particles then it is easy to see how to translate the particle hopping approach of CTAP to the spin propagation via spin-spin coupling in DSAP.  More generally, spin chains offer the possibility of creating quantum wires for solid-state quantum computers \cite{bib:Bose2007}.

\begin{figure}[tb]
\includegraphics[width=0.9\columnwidth,clip]{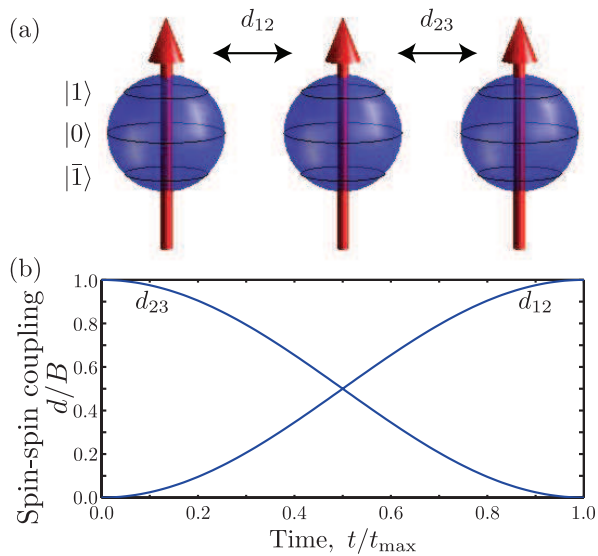}
\caption{\label{fig:3SpinSchematic} (Color online) (a) Schematic representation of the three-spin system.  Individual spin states are labelled according to their $z$ projection and spin-spin coupling is nearest-neighbour only. (b) Dark state adiabatic passage is effected by varying the couplings according to the Counter-Intuitive pulse sequence, in this case illustrated using sinusoidal modulation of $d_{12}$ and $d_{23}$.}
\end{figure}

Here we consider DSAP in a system of three spin-one particles, or equivalently qutrits depicted schematically in Fig.~\ref{fig:3SpinSchematic}(a).  We show that this system can exhibit DSAP in a fashion equivalent to that seen in spin $1/2$ systems, but it also introduces richer evolution, that is more akin to alternating adiabatic passage protocols with five states \cite{bib:ShorePRA1991,bib:PetrosyanOptComm2006,bib:JongNanotech2009}, and that observed in the Bose-Hubbard treatment \cite{bib:BradlyPRA2012}.  We discuss two methods of implementation, the first based on complete control of the three spin Hamiltonian, such as might be expected in triple dot structures, and the second using magic angle control, such as would be appropriate for dipolar coupled particles.

There has been relatively little work on quantum transport in spin-one chains, compared with that of spin-half chains, and certainly we are not aware of adiabatic passage techniques in these systems.  Understanding of the transport in spin-one chains ultimately seems to derive from the Haldane \cite{bib:Hal1983} and AKLT studies for a spin-one Heisenberg chain \cite{bib:AKLT1987}, and typically focus on the properties of the elementary excitations with the chains, e.g.  \cite{bib:Ton1970,bib:ZBK+2008,bib:CDA+2008,bib:HMN+2013}, or in some cases teleportation-based transport \cite{bib:CB2008,bib:LGMN2009,bib:GBG+2014} or entanglement swapping \cite{bib:JJ2007}.

\section{Dark state adiabatic passage with spin-one particles}\label{sect:Ideal}

Our treatment of the adiabatic passage protocol is applicable to many systems.  All that is required is three effective spin-one systems, with controllable nearest neighbour coupling.  The generic, nearest-neighbour Hamiltonian can be expressed as a function of time, $t$ as (with $\hbar = 1$)
\begin{align}
\mathcal{H} = B \sum_{i=1}^3 J_{z,i} + \left[d_{12} (t) J_1^{+}J_2^- + d_{23} (t) J_2^{+}J_3^- + \text{h.c.}\right], \label{eq:Ham}
\end{align}
where $B$ is the (possibly time varying) Zeeman energy associated with the magnetic field, $J_{z,i}$ is the spin projection operator along the $z$ axis for particle $i$, $J_i^+$ ($J_i^-$) is the spin raising (lowering) operator for particle $i$, and $d_{ij}(t)$ is the time-varying (gated) coupling energy between 
(nearest neighbour) particles $i$ and $j$.  We label the states of the particles according to their $z$ projection as $\ket{1}$, $\ket{0}$ and $\ket{\bar{1}}$.  For a given state $\psi$ we define the population in a given basis state as $P_{\alpha,\beta,\gamma} = |\braket{\alpha,\beta,\gamma}{\psi}|^2$ for $\alpha, \beta,\gamma =\bar{1},0,1$.  The passage involves ``moving" the state of a given spin, for example a 0, from particle 1 to particle 3 during the interval from $t=0$ to $t=t_{\max}$, such that at $t=0$ the system is in the state $\ket{0,\alpha,\alpha}$ and at $t=t_{\max}$ the system is in the state $\ket{\alpha,\alpha,0}$ for particular spin projections $\alpha$.  The restrictions on the allowed $\alpha$ for DSAP are discussed below.

Adiabatic passage implementation involves the Counter-Intuitive pulse ordering such that $d_{12}(0) \rightarrow 0$, $d_{23}(0) \gg d_{12}(0)$ and $d_{23}(t_{\max}) \rightarrow 0$, $d_{12}(t_{\max}) \gg d_{23}(t_{\max})$ with the $d_{jk}(t)$ smoothly varied throughout the protocol (although even this restriction is not absolute, see for example piecewise adiabatic passage \cite{bib:SMM+2007} and digital adiabatic passage \cite{bib:VG2013}).  The Counter-Intuitive pulse sequence is named for the fact that the state to be transferred is initially uncoupled whilst the non-transferred states are initially strongly coupled.  This sequence admits an infinite amount of possible implementations, and for simplicity and definiteness we choose 
\begin{align}
d_{12}(t) = d \sin^2\left(\pi t/2 t_{\max}\right), \nonumber \\ 
d_{23}(t) = d \cos^2\left(\pi t/2 t_{\max}\right),
\end{align} 
where $d$ is the maximum coupling, and the total time $t_{\max}$ is assumed long enough to ensure adiabatic evolution.  This particular sequence is shown in Fig.~\ref{fig:3SpinSchematic} (b).

We first assume that the inter-spin coupling can be directly and independently controlled.  This method is best suited to quantum dot implementations where gates can be used to independently control the exchange interaction between neighbouring spins. 

\begin{figure}[tb]
\includegraphics[width=0.9\columnwidth,clip]{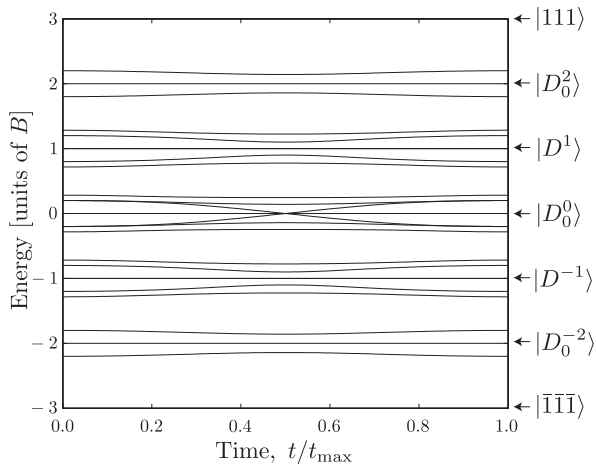}
\caption{\label{fig:Eigenspectra} Eigenspectra over the Counter-Intuitive pulse sequence with $B=1$, $d = 0.2$.  The states separate into various manifolds, which are discussed in the text.  Highlighted are some of the kets with constant energy throughout the DSAP protocol.}
\end{figure}

To gain insight into the dynamics of the three-spin system under the Counter-Intuitive pulse sequence, we present the time-dependent eigenspectra in Fig.~\ref{fig:Eigenspectra}.  We have arbitrarily set $d/B=0.2$ in the figure to separate the manifolds with different numbers of excitations.  The full solution is relatively complicated, with several degeneracies appearing, however it is easier to obtain insight into the dynamics if we focus our attention on each manifold of states centered around a given energy.  The manifolds and evolution for $E = \pm 3B$ are trivial.  These correspond to the system in $\ket{111}$ or $\ket{\bar{1}\bar{1}\bar{1}}$ states respectively, which do not respond to the coupling interaction variations and are therefore ignored in what follows.

The energy levels around $E = \pm 2B$ are relatively straightforward.  There are three states involved in each manifold.  In the $E = 2B$ manifold the basis states involved are $\ket{011}$, $\ket{101}$, and $\ket{110}$.  These show the possibility for a DSAP-like pathway where the spin state on particle one, $\ket{0}$, is  transferred to the particle three.  This can equivalently be thought of as adiabatic passage of a hole along the chain, as discussed by Benseny \textit{et al.} in the context of atomtronics \cite{bib:BensenyPRA2010}. 

\begin{figure}[tb]
\includegraphics[width=0.9\columnwidth,clip]{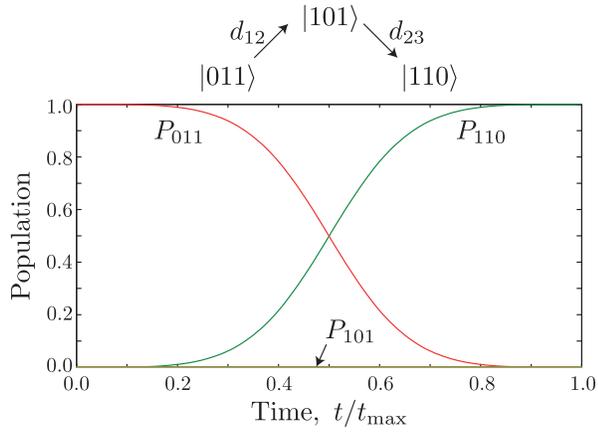}
\caption{\label{fig:DSAP} (Color online) Populations in the $E=2B$ manifold throughout the protocol determined using density matrix analysis as a function of time, confirming the DSAP evolution.  The red line is $P_{011}$ and the green line $P_{110}$.  Note that the system is initialised in the state $\ket{011}$ ($P_{011} = 1$) and evolves to the state $\ket{110}$, staying in the state $\ket{D^2_0}$ as expected, with $P_{101} = 0$ throughout the protocol.  Population in the $E=-2B$ manifold follows similarly.  For this simulation, $t_{\max} = 100~B^{-1}$. The path of the adiabatic passage is schematically shown at the top, where only the lower states are populated.  This representation also makes clear the connection between the DSAP pathway under consideration and STIRAP in the $\Lambda$ configuration.}
\end{figure}

\begin{widetext}
For our particular pulse sequence, the eigenstates around $E=2B$ are  
\begin{align}
\ket{D^{(2)}_0} &= \frac{\cos^2\left(\frac{\pi t}{2 t_{\max}}\right) \ket{011} - \sin^2\left(\frac{\pi t}{2 t_{\max}}\right) \ket{110}}{\sqrt{\cos^4\left(\frac{\pi t}{2 t_{\max}}\right) + \sin^4\left(\frac{\pi t}{2 t_{\max}}\right)}}, \\
\ket{D^{(2)}_{\pm}} &= \frac{\sin^2\left(\frac{\pi t}{2 t_{\max}}\right) \ket{011} \pm \sqrt{\frac{3 + \cos\left(\frac{2\pi t}{t_{\max}}\right)}{2}}\ket{101} - \cos^2\left(\frac{\pi t}{2 t_{\max}}\right) \ket{110}}{\sqrt{3 + \cos\left(\frac{2 \pi t}{t_{\max}}\right)}},
\end{align}
with energies
\begin{align}
E^{(2)}_0 = 2B, \quad E^{(2)}_{\pm} = 2B \pm \frac{d}{2}\sqrt{3 + \cos\left(\frac{2 \pi t}{t_{\max}}\right)}.
\end{align}
We interpret these results in the usual fashion for CTAP, namely that when $t = 0$, $d_{23} \gg d_{12}$, the system is initialized in the state $\ket{D^{(2)}_0} = \ket{011}$, and adiabatically following the Counter-Intuitive pulse sequence transfers the spin 0 state from site 1 to site 3 without modifying the spin state at site 2.  The calculated time evolution of the populations in this case is shown in Fig.~\ref{fig:DSAP}.  We note that the evolution presented here and all subsequent figures is calculated from a full solution of the time-varying Hamiltonian. When compared with the analytical results provided, full agreement is obtained. The evolution at $E = -2B$ follows from exactly the same reasoning, except that in this case the states involved are $\ket{0\bar{1}\bar{1}}$, $\ket{\bar{1}0\bar{1}}$ and $\ket{\bar{1}\bar{1}0}$.  In this case we can picture the transport as a particle moving along a chain in a CTAP process.  It should be self-evident that in these one-particle and one-hole cases, \emph{all} of the standard CTAP like results can be obtained.  In particular, extension to many-site (i.e. more than 3-site) straddling \cite{bib:GreentreePRB2004} and alternating geometries \cite{bib:PetrosyanOptComm2006,bib:JongNanotech2009} will follow trivially.  Also straightforward is the extension to the fractional protocol discussed in the context of STIRAP \cite{bib:VitanovJPB1999} and \cite{bib:Dreisow}, or adiabatic splitting in a five-site configuration \cite{bib:Chung2012}.  In the DSAP case, these splittings will produce entangled states, rather than the superpositions generated in STIRAP or CTAP, however we will not discuss these possibilities here.
\end{widetext}

The adiabaticity is a convenient way to quantify whether the system evolves along a continually varying series of connected eigenstates during evolution, or is likely to make a discontinuous jump to an unrelated eigenstate. \cite{bib:Messiah}.  Using the standard approach we parameterise the adiabaticity for any two instantaneous eigenstates $\ket{\phi_1}$ and $\ket{\phi_2}$ as
\begin{align}
A = \frac{\bra{\phi_1}\partial_t\ket{\phi_2}}{|E_{\phi_1} - E_{\phi_2}|}.
\end{align}
In particular, for the $E=2B$ manifold, we have the adiabaticity between $\ket{D^{(2)}_0}$ and either of $\ket{D^{(2)}_{\pm}}$
\begin{align}
A^{(2)} = \frac{2\sqrt{2}\pi \sin\left(\frac{\pi t}{t_{\max}}\right)}{d t_{\max}\left[3 + \cos\left(\frac{2 \pi t}{t_{\max}}\right)\right]^{3/2}}. \label{eq:Adiab2}
\end{align}

The remaining three manifolds at $E = \pm B$ and $E=0$ are not as simple due to the increase in the degeneracies.  The composition of the $E = -B$ manifold follows obviously by symmetry argument from the $E=B$ manifold, hence we do not treat it separately.

\begin{figure}[tb]
\includegraphics[width=0.9\columnwidth,clip]{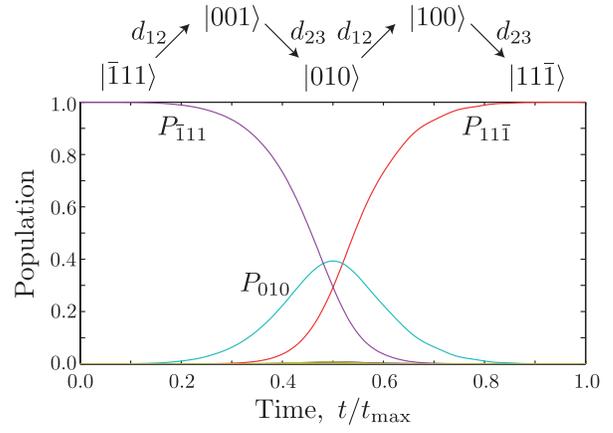}
\caption{\label{fig:E1DSAP} (Color online) Time evolution of the populations in the $E=B$ manifold when the starting state is $\ket{\bar{1}11}$, corresponding to evolution in the state $\ket{D^1}$.  The purple line shows $P_{\bar{1}11}$, the cyan line $P_{010}$ and the orange line $P_{11\bar{1}}$.  This evolution is completely analogous to the evolution observed in Alternating adiabatic passage protocols with five states.  The system is initially in the state $\ket{\bar{1}11}$ and evolves to the state $\ket{11\bar{1}}$, with transient population in the state $\ket{010}$.}
\end{figure}

The states comprising the $E=B$ manifold are in general complicated, and their form is not especially illuminating, however the states at $E = B$ exactly highlight an interesting adiabatic pathway for population transfer.  For the $E = B$ manifold, the degenerate spanning states may be taken as
\begin{align}
\ket{D^1_1} &= \frac{1}{\sqrt{3}}\left(\ket{11\bar{1}} - \ket{1\bar{1}1} + \ket{\bar{1}11}\right), \\
\text{and }\ket{D^1_2} &= \frac{\left(-d_{12}^2 + d_{23}^2\right)\ket{11\bar{1}} - d_{23}^2 \ket{1\bar{1}1} + d_{12}d_{23}\ket{010}}{\sqrt{d_{12}^4 - d_{12}^2d_{23}^2 + 2d_{23}^4}}.
\end{align}
Of course any superposition of these states is also in the $E=B$ eigenspace.  It is convenient here to take a particular superposition
\begin{align}
\ket{D^1} &\propto -\ket{D^1_2} + \frac{\sqrt{3} d_{23}^2}{\sqrt{d_{12}^4 - d_{12}^2d_{23}^2 + 2d_{23}^4}} \ket{D^1_1}, \nonumber \\
 &= \frac{d_{23}^2 \ket{\bar{1}11} - d_{12}d_{23}\ket{010} + d_{12}^2\ket{11\bar{1}}}{\sqrt{d_{12}^4 - d_{12}^2d_{23}^2 + d_{23}^4}}.
 \end{align}
A complete discussion of adiabatic evolution in degenerate subspaces can be found in the work by Rigolin and Ortiz \cite{bib:RO2012}.
The state $\ket{D^1}$ is analogous to the states found with Alternating STIRAP \cite{bib:ShorePRA1991} and CTAP protocols with five sites (ACTAP$_5$)\cite{bib:PetrosyanOptComm2006,bib:JongNanotech2009}.  
In the adiabatic limit with Counter-Intuitive pulse ordering, the passage is from $\ket{\bar{1}11}$ to $\ket{11\bar{1}}$.  This can be understood as transport of the $\ket{\bar{1}}$ state from spin 1 to spin 3 via the states $\ket{001}$, $\ket{010}$, and $\ket{100}$, with the populations $P_{001} = P_{100} = 0$, and transient population in state $\ket{010}$.  This evolution is shown in Fig.~\ref{fig:E1DSAP}.

The manifold of seven states around $E=0$ is also very interesting.  At time $t = t_{\max}/2$ (i.e. when $d_{12}=d_{23}$) there is a clear anti-crossing arising from the adiabatic passage transfer, and also a true three-state crossing.  The state 
\begin{align}
\ket{D^0_0} = \frac{d_{23}\left(\ket{01\bar{1}} - \ket{0\bar{1}1}\right) - d_{12}\left(\ket{1\bar{1}0} - \ket{\bar{1}10}\right)}{\sqrt{2}\sqrt{d_{12}^2 + d_{23}^2}},
\end{align}
remains an eigenstate with $E=0$ throughout the protocol and demonstrates adiabatic passage of the spin 0 state from site 1 to site 3,  with the rest of the chain in a particular entangled state.

\section{Qutrit Transport Protocol}

The DSAP protocols involving $\ket{011}$ and $\ket{\bar{1}11}$ are quite similar, and can both be effected by the same gate control sequence, i.e. the same variation in the $d_{ij}(t)$.  Although the control sequence is the same in each case, the properties of the evolution differ quantitatively.   The protocol involving $\ket{\bar{1}11}$ is slightly less adiabatic than the $\ket{011}$ protocol,  which follows from the form of the null states, as discussed in Ref.~\cite{bib:JongNanotech2009}.  We compare the rate limiting adiabaticities for the two evolutions in the $E = \pm 2 B$ and the $E=\pm B$ manifolds, for the transitions between the states $\ket{D_0^{(\pm2}}$ and $\ket{D_{\pm}^{\pm2}}$, denoted $A^{(2)}$, with the adiabaticity between the states $\ket{D_0^{\pm 1}}$ and $\ket{D_{\pm}^{\pm 1}}$ denoted $A^{(1)}$. One needs to be careful about applying the adiabatic theorem within  degenerate subspaces. Adiabaticities involving degenerate subspaces are taken relative to the closest states outside the degenerate manifold.  Under these conditions we find that
\begin{align}
A^{(2)}\left(t = \frac{t_{\max}}{2}\right) &= \frac{\pi}{t_{\max}d}, \\
A^{(1)}\left(t = \frac{t_{\max}}{2}\right) &= \frac{2\pi}{t_{\max}d}\sqrt{\frac{3}{7}\left(4 + \sqrt{2}\right)}.
\end{align}
The presence of parallel DSAP channels in the same system suggests two interesting corollaries.  Firstly, this DSAP protocol would allow the adiabatic transport of a qutrit encoded in one of the spins, i.e. where the initial state of the chain is a superposition $\alpha\ket{111} + \beta\ket{011} + \gamma \ket{\bar{1}11}$.  As $A^{(1)}>A^{(2)}$, for high fidelity qutrit transport, the \emph{worst case} adiabaticity must be used to ensure adiabatic passage for the qutrit as a whole.  Note that this is an advantage of adiabatic passage, as a non-adiabatic scheme would require gate operations of precise durations, such that equal populations were transferred from each of the starting states, which is more restrictive than simply requiring high fidelity population transfer for the states independently. Secondly, we can see that an error that only affects the first spin will not be communicated to the rest of the chain.  Although this latter point is appealing for the purposes of quantum information transfer, it is clear that the converse is \emph{not} true, and in general, errors in the chain do affect the transport protocol.

The configuration described above, where the non-data qutrits are in the state $\ket{11}$ is not the only possible state to allow qutrit transport via DSAP.  By examining the null states described above, we observe that complete qutrit transport can be achieved when the two non-data qutrits are in the states $\ket{11}$, $\ket{\bar{1}\bar{1}}$ and $(1/\sqrt{2})(\ket{1\bar{1}} - \ket{\bar{1}1})$.  Also, any superposition of these states of the chain will allow for DSAP transport, including entangled states of the form $\sin\varphi\ket{11} - \cos\varphi\ket{\bar{1}\bar{1}}$ for arbitrary $\varphi$, although we note that the state $(1/\sqrt{2})(\ket{1\bar{1}} + \ket{\bar{1}1})$ does \emph{not} allow qutrit transport.

\section{Dipole coupling effected via magic-angle control}\label{sect:dipole}

Not all spin systems have obvious mechanisms to allow independent control of the spin-spin coupling via some gate mechanism.  Magic angle coupling can be used for controllable dipole-dipole coupling to effect the desired Counter-Intuitive pulse sequence, a mechanism proposed for controlled coupling in a dipolar phosphorus in silicon quantum computer \cite{bib:SousaPRA2004}.  The approach here is to vary the magnetic field direction along a trajectory that zeros the coupling between spins 1 and 3, whilst varying $d_{12}$ and $d_{23}$ according to the Counter-Intuitive pulse sequence.

The dipole-dipole coupling between two spins, $j$ and $k$, in a magnetic field is ($\hbar = 1$)
\begin{align}
d_{jk} (\theta_{jk},r_{jk}) = \frac{\gamma_j \gamma_k}{r_{jk}^3} \left(3 \cos^2 \theta_{jk} - 1\right),
\end{align}
where the $\gamma$ are the dipole moments, $\omega = \gamma B$, $\theta_{jk}$ is the angle between the magnetic field and the line joining the spins, and $r_{jk}$ is the separation between spins.  Now when $|\cos \theta_{jk}| = 1/\sqrt{3}$ we must have $d_{jk} = 0$, hence there is no coupling.

\begin{figure}[tb!]
\includegraphics[width=0.9\columnwidth,clip]{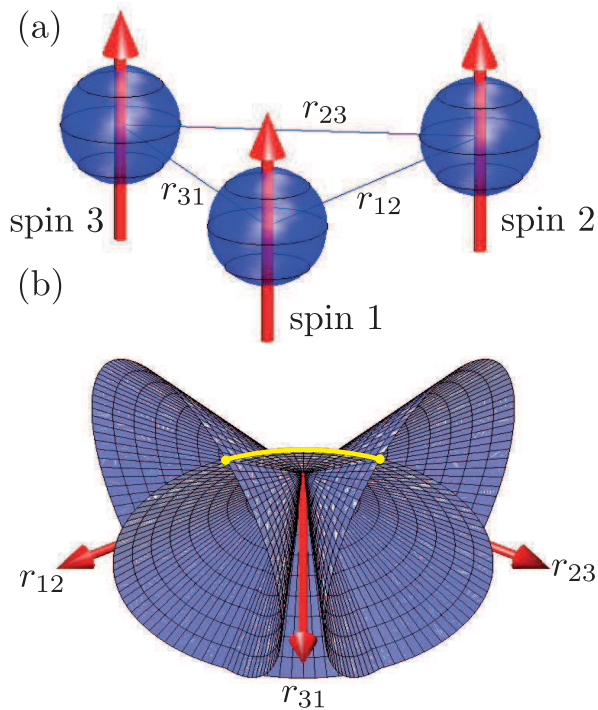}
\caption{\label{fig:DipoleCoupling} (Color online) (a) Arrangement of three spins in the $x-y$ plane. (b) Inter-spin separations with magic angles for each pair of spins marked.  Because there are points of intersection of the magic angles, it is possible to define a magnetic field trajectory that effects the Counter-Intuitive pulse sequence, and one such trajectory is shown here in yellow with the start/stop points marked with yellow dots.}
\end{figure}

To demonstrate the appropriate control of the dipole-dipole coupling, we consider an arrangement of three equally spaced spins in the $x-y$ plane, as shown in Fig.~\ref{fig:DipoleCoupling}(a).  To understand an implementation of the Counter-Intuitive pulse sequence for DSAP from spin 1 to spin 3, in Fig.~\ref{fig:DipoleCoupling}(b) we show the rays between the spins and the cones show the magic angles for the magnetic field to null the dipolar coupling.  A possible Counter-Intuitive pulse sequence trajectory is highlighted in yellow, where the central cone corresponds to nulling the 1-3 coupling and the end points (yellow dots) correspond to the case where either the 1-2 or 2-3 coupling is also cancelled.  The magic magnetic trajectory defined by this configuration is
\begin{align}
\mathbf{B} = B \left[ \cos \varphi(t), \cot \theta_m, \sin \varphi(t) \right], \label{eq:MagicAngle}
\end{align}
where $\theta_m$ is the magic angle and $\varphi(t)$ specifices the time-varying trajectory of the Counter-Intuitive pulse  sequence and 
\begin{align}
\cos \left[\varphi(0)\right] &= \pi - \frac{\cot \theta_m \sin (2\pi/3)}{\cos(2\pi/3) - 1}, \\
\cos \left[\varphi(t_{\max})\right] &= \frac{\cot \theta_m \sin (2\pi/3)}{\cos(2\pi/3) - 1}.
\end{align}

\begin{figure}[tb]
\includegraphics[width=0.9\columnwidth,clip]{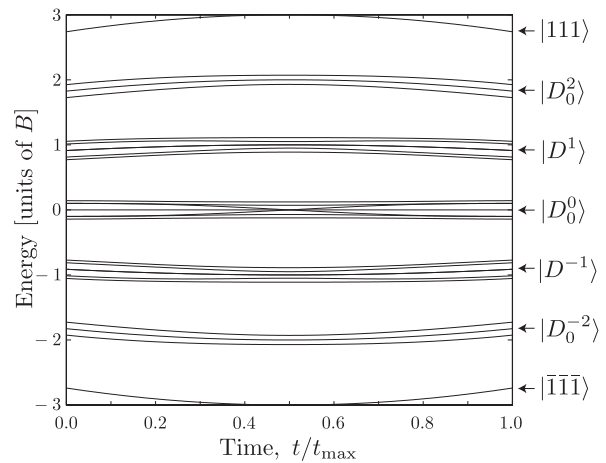}
\caption{\label{fig:DipoleEvals} Eigenspectrum for the magic angle control protocol, where the angle of the applied field is according to the trajectory outlined in Eq.~\ref{eq:MagicAngle} for the case that the three spins are located at the vertices of an equilateral triangle.}
\end{figure}

Following a magic angle trajectory of the form envisaged here perforce changes $B_z$ as well as the $d_{jk}$, and hence the eigenspectrum, shown in Fig.~\ref{fig:DipoleEvals} is slightly more complicated than the simpler case studied in Sect.~\ref{sect:Ideal}.  Nevertheless, the overall structure of the manifolds is unchanged from our earlier treatment, with the trajectories appearing to `bend' due to the varying $z$ component of the magnetic field relative to the dipoles.  However the relative ordering of the states and their degeneracy is unaffected by using this control scheme, rather than the earlier more idealised approach where the magnitude of $B_z$ is constant.  We note that the form of the dipole coupling ensures square sinusoidal variation in the coupling coefficients as we assumed for the `ideal' version.  

\section{Possible Experimental Realisations} \label{sect:Expt}
There are many possible systems in which the spin-one version of DSAP could be implemented. Here we briefly explore two such platforms, spin-based quantum dot arrays and spin-one defects in diamond. We also note other possibilities to be explored for implementation, such as  chains of trapped ions \cite{bib:MT2005}, NiCl$_2$-4SC(NH$_2$)$_2$ \cite{bib:ZBK+2008,bib:CDA+2008}, and liquid-phase NMR, e.g.  via deuterated molecules \cite{bib:DMKK2003,bib:Jon1011}.

\subsection{Quantum dots}
The design and engineering of controlled electrostatic potential landscapes in two-dimensional electron gasses has led to the production of quantum dots with remarkable and beautiful quantum properties.  A single quantum dot can typically be manipulated to hold a pre-set number of electrons, and the singlet-triplet subspace of a two-electron quantum dot has been identified as a useful qubit encoding \cite{bib:PCM+2005}.  Conversely, the triplet subspace defines a spin-one subspace that may be used for our purposes. Triple dots have been demonstrated several times \cite{bib:SGG+2007,bib:GGK+2012}, although we are not aware of any that have specifically operated in the particular six-electron configuration necessary to test spin-one DSAP.  Entanglement between pairs of double-dots operating in the single-triplet basis have also been demonstrated \cite{bib:SDH+2012}.  This latter work shows controlled coupling similar to that required to test DSAP, with two-qubit coupling times of order 100~ns.

\subsection{Nitrogen-vacancy centre in diamond}
The negatively-charged nitrogen-vacancy (NV) centre in diamond has emerged as an extremely interesting system for room-temperature quantum information processing.  This is because the ground state spin levels, which form a natural spin-one system, are long lived at room temperature and they can be optically initialsed and read out at room temperature.  However, most of the concepts for scalable quantum computing with NV centres require cryogenic temperatures due to spectral broadening of the main optical transitions \cite{bib:GFM+2008}.  If there were no requirements for coherent coupling from the ground state manifold to the excited state, then it may be possible to construct a room-temperature quantum computer based on NV, and this is the subject of the proposal by Yao \textit{et al.} \cite{bib:YJG+2012}.

Whilst a deterministically created array of three NV centres at these separations has not been achieved, pair implantations (i.e. implantation of N$_2^+$) have been used to create coupled NV-N systems \cite{bib:GDP+2006} and NV-NV systems \cite{bib:NKN+2010} have also been formed by implantation through a mask.  It should be possible to extend these methods to create small clusters of implanted N, which could be searched to identify a cluster of three NV centres.    Ref.~\cite{bib:NKN+2010} demonstrated dipole-dipole coupling between the NV centres that were around 10~nm apart.  The techniques outlined in Sect.~\ref{sect:dipole} should enable a three-NV complex to perform spin-one DSAP.  Other fabrication techniques that have the required precision include low-energy nano implantation through a nano stencil \cite{PWM+2010} and ultra cold ion source implanters \cite{MVB+2006}.

An alternative to explore electron-spin coupling would be to look at the nuclear spin coupling in the three N system.  The $^{14}$N nucleus is also a spin-one particle \cite{bib:HBF1993}.  In fact, Bermudez \textit{et al.} have already proposed a two-qubit operation between two N nuclear spins in diamond, mediated by the electron spin \cite{bib:BJP+2011}.

\section{Conclusions}

We have shown that the concept of dark-state adiabatic passage (DSAP) \cite{bib:Ohshima2007} can be extended from spin-half particles to arrays of spin-one particles.  In particular, we have shown adiabatic pathways for an array of three spin one particles either direct control of the qutrit-qutrit coupling, or by alignment control of a uniform, external DC magnetic field.  

In the case of conventional DSAP, the state of a single qubit is transmitted along a chain of qubits using the Counter-Intuitive pulse sequence.  The canonical example, where the chain qubits are all either aligned parallel or anti-parallel to the quantisation axis, is formally equivalent to the case of Coherent Tunneling Adiabatic passage of either particles \cite{bib:EckertPRA2004,bib:GreentreePRB2004} or holes \cite{bib:BensenyPRA2010}.  The spin-one version of DSAP is certainly richer than the spin-half or pseudo spin-half version.  We have shown qutrit transport across three spin-one particles when the other two spins are in one of the states $\ket{11}$, $\ket{\bar{1}\bar{1}}$ or $\ket{1\bar{1}} - \ket{\bar{1}1}$.  

Whilst the transport of the qutrit is adiabatically protected, it is important to stress that single qutrit errors on the chain particles (i.e. the non-data qutrits) will in general cause errors in the protocol.   
One may think of the error as producing another effective particle, and then particle-particle interactions will become important and will likely destroy the desired or predicted transport outcome.  The sensitivity of the intended spin passage to errors in the non-data qutrits  appears to be a property of most bus-type proposals for quantum information transport if a defined propagation direction is not maintained.

\section*{Acknowledgements}
ADG would like to thank Jan Jeske, Jared Cole and Andrew Martin for useful conversations. ADG also acknowledges the ARC for financial support (DP130104381).  BK was partially supported by the Brazilian agencies FAPERJ,  CNPq. Work performed as part of the Brazilian National Institute for Science and Technology on Quantum Information.


\end{document}